\documentclass[conference, 10pt, onecolumn]{IEEEtran}

\ifCLASSINFOpdf

\else

\fi

\usepackage{graphicx}
\usepackage{color}
\usepackage{placeins}
\usepackage{float}
\usepackage{hyperref}
\usepackage{tabularx,colortbl,subcaption}

\begin{document}
%
\title{Compact 60GHz On-Chip Antenna in 65nm CMOS Technology With Circular and Linear Polarizations for Millimeter-Wave Applications}

\author{\IEEEauthorblockN{Arash Masrouri, Nasrin Amiri}
\IEEEauthorblockA{Electrical Engineering Department\\
Islamic Azad University, South Tehran Branch\\
Tehran, Iran\\
st\_a\_masrouri@azad.ac.ir, n\_amiri@azad.ac.ir\\}
}
\maketitle

\begin{abstract}
In this paper, design and simulation of a compact on-chip antenna for 60 GHz band applications is presented. Antenna is designed in 65 nm TSMC technology. This technology consists of 9 metal layers. A 0.7 mm$\times$0.7 mm patch with slots is located on 9th layer and first layer is used as ground plane. Total chip size is 0.8 mm$\times$0.8 mm. In order to enhance radiation efficiency and control axial ratio, four slots are adjusted on ground plane. Circular and linear polarizations are achievable by proper port excitation. Maximum efficiency of the antenna is about 60$\%$ at 59.6 GHz. Simulated antenna bandwidth ranges from 58.7 GHz as band start and 62.3 GHz as band stop. Peak gain values for circularly and linearly polarized modes are -0.6 dB and -0.9 dB at 0$^{\circ}$ respectively.
\end{abstract}

\IEEEpeerreviewmaketitle

\section{Introduction}
Tendency to broadband, short-range, wireless personal communications has driven network designers toward millimeter-wave technology to raise market requirements. Approved unlicensed 60 GHz band with 7 GHz available spectrum is a good candidate to facilitate the need of higher data rates and wider bandwidth. IEEE 802.15.3c as a pioneer standard of Giga-bit-per-second data transfer rate also proposes four 2.16 GHz channels for users in application layer \cite{sun201360}, \cite{fisher200760}.

According to high attenuation characteristics of propagation medium around propose spectrum for 60 GHz, this frequency band is a proper choice for distances in order of a room and indoor communications \cite{fakharzadeh2016compact}. Imaging, HD video transfer and unmanned vehicles are just some applications of the proposed frequency band.

Cabling, wiring and interconnections between radios and other communication blocks such as LNA and PA, lead to signal level descent. Antenna integration can overcome this problem. Here appear two solutions: Antenna-on-Chip and Antenna-in-Package. On-chip antennas over CMOS technology propose several advantages which lead to low signal loss, higher integration levels and cost reduction in manufacturing process \cite{chuang201160}. 

To consider quality of transmitted or received signal, polarization becomes and important issue. CP antennas can reduce signal degradations in multipath propagation environments and show less sensitivity to antenna rotation \cite{manabe1996effects}, \cite{fakharzadeh2013integrated} 

In this paper, authors describe a compact, On-chip antenna with circular and linear polarizations in 65 nm TSMC technology for 60 GHz band applications over a 270 $\mu$m silicon substrate. Overall chip size is 0.8 mm$\times$0.8 mm.

\begin{figure}[H]
	\begin{center}
		\includegraphics[width=3in]{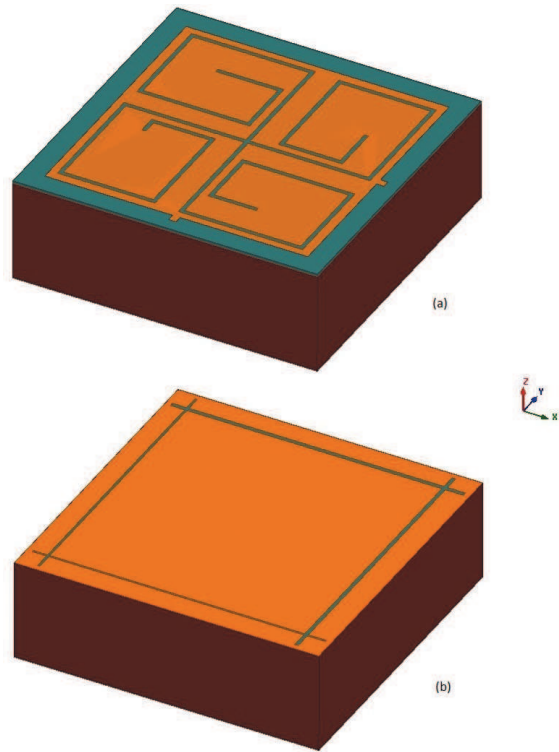}
		\caption{Antenna Model: (a) Patch (b) Ground Plane }\label{fig1}
	\end{center} 
\end{figure}
\section{Antenna Design}

Proposed antenna is realized on 65 nm CMOS process. This technology consists of 9 metal layers. Patch is designed on 9th layer. Slots are adjusted to get desire resonance frequency. The patch size is 0.7 mm$\times$0.7 mm. First metal layer is used as ground plane. Distance between patch and ground plane is about 5 $\mu$m. total chip size is 0.8 mm$\times$0.8 mm. Silicon substrate has 270 $\mu$m height. The antenna provides circular and linear polarizations based upon dual or single port excitation. Most of on-chip antennas suffer from low gain levels due to loss in silicon substrate and limited design area. Four slots are prepared on ground plane to enhance antenna efficiency. These slots also control axial ratio to get circular polarization. Position of these slots are crucial parameters in design process. Fig.~\ref{fig1} illustrates proposed antenna model.

\section{Simulation Results}
\begin{figure}[h!]
	\begin{center}
		\includegraphics[width=3.5in]{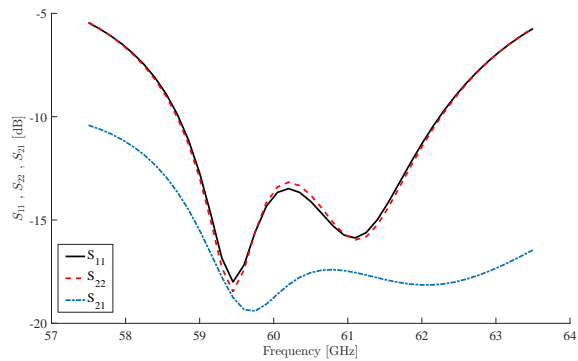}
		\caption{S parameters of the proposed antenna}\label{fig2}
	\end{center}
\end{figure}

Fig.~\ref{fig2} shows simulated S-parameters of the proposed antenna. Bandwidth measured from -10 dB is about 3.6 GHz, from 58.7 GHz to 62.3 GHz. S21 also confirms good isolation between ports. While two ports are excited with same amplitude and 90$^{\circ}$ phase difference, antenna has axial ratio of below 3dB along with simulated bandwidth.

\begin{figure}[h!]
	\begin{center}
		\includegraphics[width=3.5in]{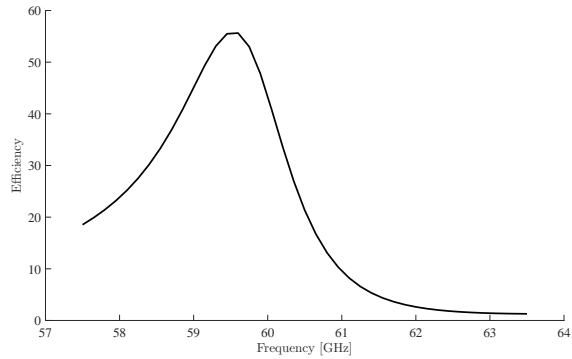}
		\caption{Radiation efficiency of the proposed antenna}\label{fig3}
	\end{center}
\end{figure}

In Fig.~\ref{fig3} radiation efficiency of the proposed antenna for dual and single port operation modes is depicted. Along with second channel of the 60 GHz band according to 802.15.3c channelization, the antenna efficinecy rises to its maximum of 60$\%$. For 60 GHz this value is 45$\%$, Then the efficiency drops along channel band stop frequency. 

\begin{figure}[H]
	\centering
	\begin{subfigure}[b]{0.5\textwidth}
		\centering
		\includegraphics[width=3in]{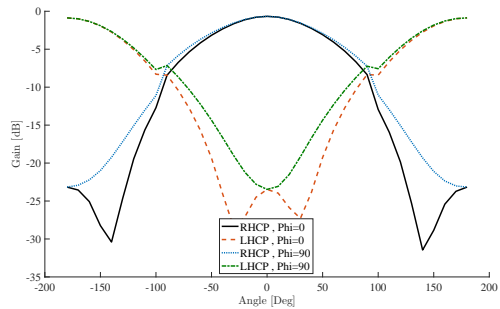}
		\caption{Gain of the proposed antenna for CP mode}
		\label{fig4}
	\end{subfigure}
	\begin{subfigure}[b]{0.5\textwidth}
		\centering
		\includegraphics[width=3in]{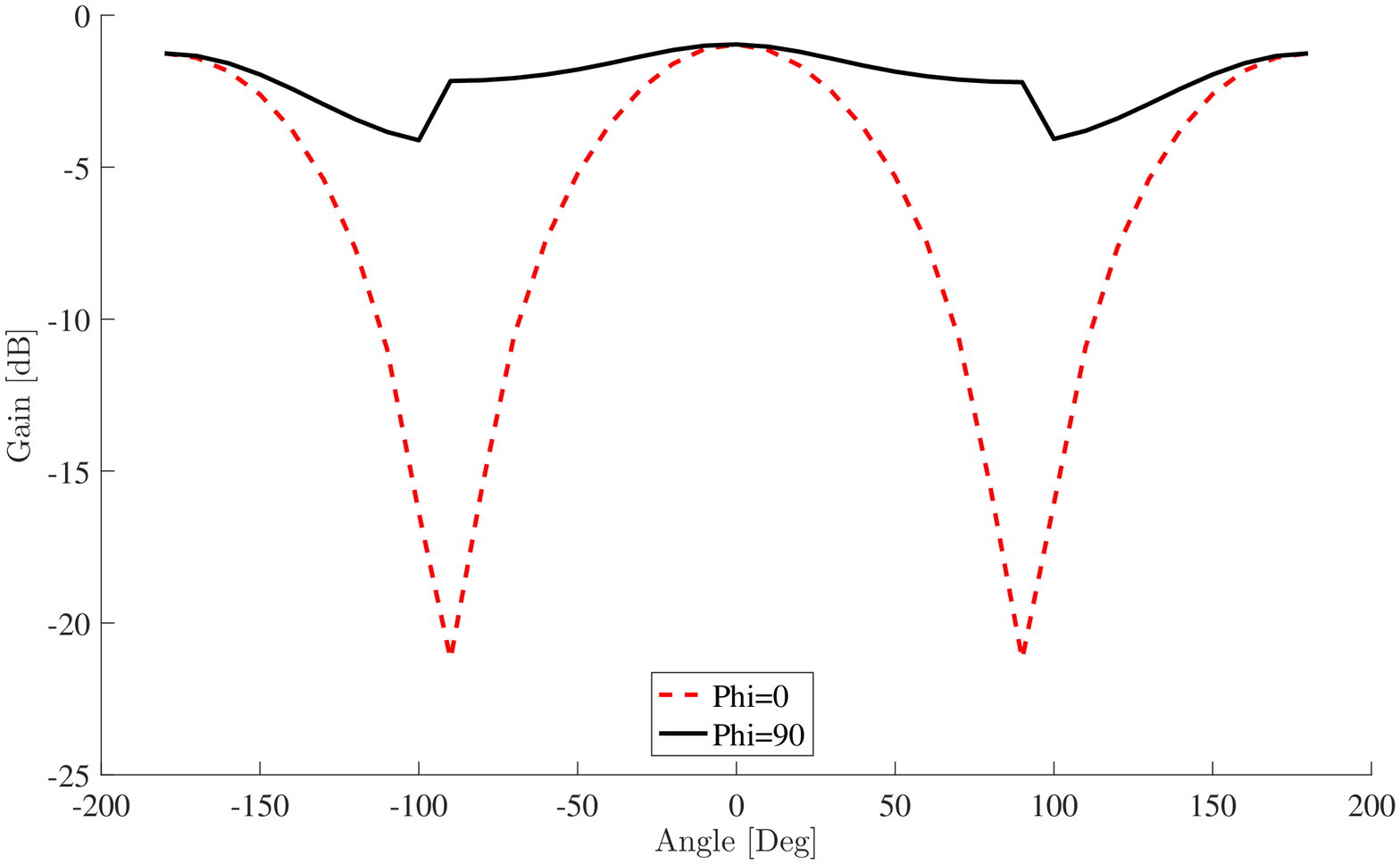}
		\caption{Gain of the proposed antenna for Linearly polarized mode}
		\label{fig5}
	\end{subfigure}
	\caption{Gain of the proposed antenna}
	\label{figtot}
\end{figure}

Gain values of the proposed antenna are also shown for both principal planes in Fig.~\ref{fig4} and Fig.~\ref{fig5}. According to the charts, peak gain value is -0.6 dB at 0$^{\circ}$ for CP mode and -0.9 dB at 0$^{\circ}$ for single port excitation.
Beamwidth of the antenna is 110$^{\circ}$ and 118$^{\circ}$ for RH circular polarization at Phi=0 and Phi=90 planes, respectively. 
%

\end{document}